\documentclass[a4paper,fleqn,usenatbib]{mnras}

\usepackage{newtxtext,newtxmath}

\usepackage[T1]{fontenc}
\usepackage{ae,aecompl}

\usepackage{graphicx}	\usepackage{amsmath}	\usepackage{amssymb}	\usepackage{caption}
\usepackage{subcaption}
\usepackage{hyperref}

\usepackage[english]{babel}
\usepackage[autostyle]{csquotes}
\MakeOuterQuote{"}

\usepackage{color}

\newcommand{\Msun}{M$_{\odot}$}	\newcommand{\kms}{kms$^{-1}$}
\newcommand{\Reff}{R$_{\mathrm{e}}$}
\newcommand{\ppxf}{\texttt{pPXF}}

\title[A MUSE view of the low-mass IMF in NGC 1399]{The stellar population and initial mass function of NGC\,1399 with MUSE}

\author[S. P. Vaughan et al.]{Sam P. Vaughan,$^{1}$\thanks{E-mail: sam.vaughan@physics.ox.ac.uk (SPV)}, Roger L. Davies$^{1}$,  Simon Zieleniewski$^{1}$
\newauthor{and Ryan C. W. Houghton$^{1}$}
\\
$^{1}$Department of Astrophysics, University of Oxford, Denys Wilkinson Building, Keble Road, Oxford, OX1 4RH, UK\\
}

\date{Accepted 2018 May 30. Received 2018 May 30; in original form 2018 April 12}

\pubyear{2018}

\begin{document}
\label{firstpage}
\pagerange{\pageref{firstpage}--\pageref{lastpage}}
\maketitle

\begin{abstract}

We present spatially resolved measurements of the stellar initial mass function (IMF) in NGC\,1399, the largest elliptical galaxy in the Fornax Cluster. Using data from the Multi Unit Spectroscopic Explorer (MUSE) and updated state-of-the-art stellar population synthesis models from \cite{2018ApJ...854..139C}, we use full spectral fitting to measure the low-mass IMF, as well as a number of individual elemental abundances, as a function of radius in this object. We find that the IMF in NGC\,1399 is heavier than the Milky Way in its centre and remains radially constant at a super-salpeter slope out to 0.7 \Reff. At radii larger than this, the IMF slope decreases to become marginally consistent with a Milky Way IMF just beyond \Reff. The inferred central V-band M/L ratio is in excellent agreement with the previously reported dynamical M/L measurement from \cite{2006MNRAS.367....2H}. The measured radial form of the M/L ratio may be evidence for a two-phase formation in this object, with the central regions forming differently to the outskirts. We also report measurements of a spatially resolved filament of ionised gas extending 4\arcsec\,(404 pc at $D_{\mathrm{L}}=21.1$ Mpc) from the centre of NGC\,1399, with very narrow equivalent width and low velocity dispersion ($65\pm14$ \kms). The location of the emission, combined with an analysis of the emission line ratios, leads us to conclude that NGC\,1399's AGN is the source of ionising radiation. 
\end{abstract}

\begin{keywords}
galaxies: stellar content -- galaxies: elliptical and lenticular, cD -- galaxies: individual -- galaxies: abundances -- galaxies: formation
\end{keywords}

\section{Introduction}

The stellar Initial Mass Function (IMF) is a key ingredient in our attempts to model and understand the evolution of galaxies. The IMF describes the number density of stars on the zero-age main sequence in a population as a function of stellar mass. It is well known that the lifetime, colour and eventual fate of a single star are determined by its mass, and as a result a collection of stars with different proportions of low-mass stars to high-mass stars (i.e a different IMF) can appear markedly different.

Whilst the detailed process of star formation is complicated and technically challenging to simulate and model \citep[see e.g. ][for a review]{2007ARA&A..45..565M}, the form of IMF can be described by a simple functional form. \cite{1955ApJ...121..161S} undertook a stellar census of the Milky Way and measured the IMF to be a single power law between 0.4 \Msun\,and 10 \Msun; $\xi(m)=Cm^{-x}$, with slope of $x=2.3$. Current popular IMF models often use this slope at masses greater than 1 \Msun (and extend the high-mass cut-off to be $\geq100$ \Msun) but follow a lognormal distribution \citep{2003PASP..115..763C} or a broken power law \cite{2002Sci...295...82K} at masses below 1 \Msun.

A longstanding assumption in extragalactic astronomy has been that the IMF is universal; that is, constant across cosmic time and invariant between (and within) galaxies. This was mainly motivated by the fact that, until recently, measuring the IMF in unresolved stellar populations (where direct star counts are infeasible) was a technically very challenging, if not impossible, task. To  infer the IMF in an unresolved population spectroscopically, one has to measure the contribution of low-mass stars to the overall galaxy spectrum. This is both observationally difficult, due to the fact that stars below 0.4 \Msun\,contribute on the order of $1$\% of a galaxy's light for a Milky Way IMF \citep{2012ApJ...747...69C}, but also suffers from an important degeneracy; at low resolution, red giants and low-mass ($<0.5$ \Msun) dwarfs at the same effective temperature have very similar spectra, but the giants are orders of magnitude brighter. For example, M giants may have a luminosity of $10^{4}$ L$_{\odot}$ \citep[e.g.][]{2015RAA....15.1154Z} whilst an mid M-dwarf may be only 0.01 L$_{\odot}$ \citep{2009ApJ...698..519K}. It is therefore difficult to tell whether a spectrum contains light from a large number of M-dwarf stars, implying a steep low-mass IMF, or just a small number of M-giants, leading to a more conventional mass function.

The discovery of gravity sensitive absorption indices in the spectra of low-mass stars \citep[e.g.][]{1945ApJ...101..265K,1969PASP...81..527W,1977ApJ...211..527W} led to a way to break this degeneracy. Certain spectral features change strength depending on the surface gravity of the atmosphere they reside in, implying that they vary between dwarf and giant stars. Measuring the strength of these features in the unresolved spectra of a population can tell us its dwarf-to-giant ratio, and hence the IMF. 

This method was pioneered by \cite{1962ApJ...135..715S} in a number of early-type galaxies (ETGs), with many later studies of the dwarf-star population in M31 using similar techniques \citep{1971ApJS...22..445S, 1978ApJ...221..788C, 1980ApJ...235..405F}. Research in this area continued in this fashion on a number of objects \citep[e.g.][]{1986ApJ...311..637C, 1993ApJ...406..142C, 2003MNRAS.339L..12C} until work by \cite{2010Natur.468..940V} led to renewed interest in the topic. 

Recent advancements in the modelling of stellar populations \citep[e.g.][]{2012ApJ...747...69C, 2016MNRAS.463.3409V, 2018ApJ...854..139C} have led to a number of studies measuring a steepening of the low-mass IMF slope with a galaxy's central velocity dispersion and/or stellar mass \citep[e.g.][]{2012ApJ...760...71C, 2013MNRAS.429L..15F, 2013MNRAS.433.3017L,2015ApJ...803...87S}. Furthermore, studies of the variation of the IMF within galaxies \citep[][hereafter S18]{2015MNRAS.447.1033M,2015MNRAS.451.1081M,2017ApJ...841...68V,2017ApJ...837..166C,2018MNRAS.477.3954P,2018MNRAS.478.4084S} have built up the picture of this bottom-heavy IMF (i.e. an excess of dwarf stars) being concentrated in the nuclei of massive ETGs, with Milky-Way like IMFs measured in the outskirts of these objects \citep[although see][who find no significant radial IMF gradient on average in their sample of seven nearby ETGs]{2018MNRAS.tmp.1180A}. 

Such a picture ties in neatly with the "two-stage" formation scenario for ETGs. This hypothesises that the cores of elliptical galaxies began life as a compact central starburst which grows via smooth gas accretion and dissipative mergers, whilst their stars at larger radii were accumulated through minor mergers of smaller satellite objects \citep[e.g.][and references therein]{2013IAUS..295..340N}. This implies that the stars currently comprising the outskirts of nearby ETGs formed in less massive halos and different physical conditions to those in the core, which may naturally explain the presence of spatial gradients in the IMF. 

However, whilst IMF determinations from measuring the integrated light of ETGs and a number of dynamical methods (e.g strong-lensing measurements and modelling the gravitational potential of galaxies using their kinematics) are broadly in agreement, there are examples of these techniques giving different results on a galaxy by galaxy basis \citep{2014MNRAS.443L..69S}. Whilst a likely reason for this discrepancy is an aperture difference between the dynamical and spectroscopic studies \citep{2016MNRAS.463.3220L}, \cite{2017ApJ...845..157N} present a similar tension in two nearby ETGs where such aperture differences are negligible. They find that the stellar M/L derived from lensing and dynamical methods is different from that inferred from fitting the integrated light at the 2-3$\sigma$ level, although using more flexible stellar population synthesis models (with a variable low-mass cut-off, or an entirely non-parameteric IMF, neither of which we employ in this work) and a invoking a reduction in the central dark-matter fraction of the ETGs reduces this tension to below $1\sigma$.

Massive ETGs with Milky-Way like (rather than bottom heavy) IMF normalisations have also been found \citep[e.g.][]{2015MNRAS.449.3441S}, and some studies invoke abundance variations, rather than a change in the IMF, to explain their spectral measurements of massive ETGs (e.g \citealt{2015MNRAS.452..597Z, 2016ApJ...821...39M, 2017MNRAS.465..192Z, 2017MNRAS.468.1594A}, and see \citealt{2018MNRAS.475.1073V} for an example of the difficulties in disentangling the two effects). There may also be a question concerning whether the effects of a variable contribution from asymptotic giant branch stars or a change in the low-mass cut off (LMCO) of star formation may mimic changes in gravity sensitive indices attributed to a change in the IMF slope \citep{2015MNRAS.453.4431T}, although \cite{2017ApJ...837..166C} study a non-parametric form of the IMF where the LMCO is a free parameter in their modelling and still recover a bottom-heavy population of stars. 

A number of these discrepancies may be attributed and explained by intrinsic scatter in the IMF of ETGs, and so building up a large sample of galaxies with state-of-the-art IMF determinations is important. To this end, we present resolved IMF measurements of NGC\,1399, the largest ETG in the Fornax Cluster. 

As one the largest nearby massive galaxies, NGC\,1399 is well studied in the literature. It has the core-like surface brightness profile characteristic of a supergiant elliptical \citep{1986ApJS...60..603S, 2017A&A...603A..38S}, hosts two jets of radio emission and a pair of X-ray cavities \citep{1988ApJ...325..180K, 2008MNRAS.383..923S, 2017ApJ...847...94S} and has a been the subject of a number of dynamical studies in the optical \citep[e.g.][]{1989ApJ...336..639B, 1989ApJ...344..613F, 1994A&AS..105..433L, 1998A&AS..133..325G, 2000AJ....119..153S, 2007ApJ...671.1321G} and the near infrared \citep{2006MNRAS.367....2H, 2008A&A...485..425L}. 

We use full spectral fitting with the SPS library of  \cite{2018ApJ...854..139C} to determine the low-mass IMF slope, as well as a number of other stellar population parameters, as a function of radius in this galaxy. In section \ref{sec:obs} we present our observations and discuss how the data were reduced. In section \ref{sec:spectral_fitting} we describe our methodology, and present results in \ref{sec:results}. We draw our conclusions in \ref{sec:conclusions}.

In this work, we define the IMF to be a three part power law as follows:

\begin{equation}
    \xi(m)=\left\{
                \begin{array}{ll}
                  C_{1}m^{-x_1}  \mathrm{\,for\,}  0.08\mathrm{M}_{\odot} \leq m < 0.5\mathrm{M}_{\odot} \\
                  C_{2}m^{-x_2}  \mathrm{\,for\,}  0.5\mathrm{M}_{\odot} \leq m < 1\mathrm{M}_{\odot}\\
                  C_{3}m^{-x_3} \mathrm{\,for\,}  m \geq 1\mathrm{M}_{\odot}
                \end{array}
              \right.
\end{equation}

The normalising constants $C_{i}$ are chosen to ensure the IMF is continuous at each $m$. In all subsequent analysis, we fix the high-mass slope $x_3$ to be 2.3. We use the phrase "Milky Way IMF" to refer to the IMF from \cite{2002Sci...295...82K}, which has $x_3=x_2=2.3$ and $x_1$=1.3. We assume a Planck cosmology of $H_{0}=67.8$ kms$^{-1}$Mpc$^{-1}$ and $\Omega_{\mathrm{m}}=0.308$ \citep{2016A&A...594A..13P} to give the luminosity distance of NGC\,1399 as 21.1 Mpc.

\section{Observations and Data Reduction}
\label{sec:obs}

\begin{figure*}
\begin{center}
\includegraphics[width=\linewidth]{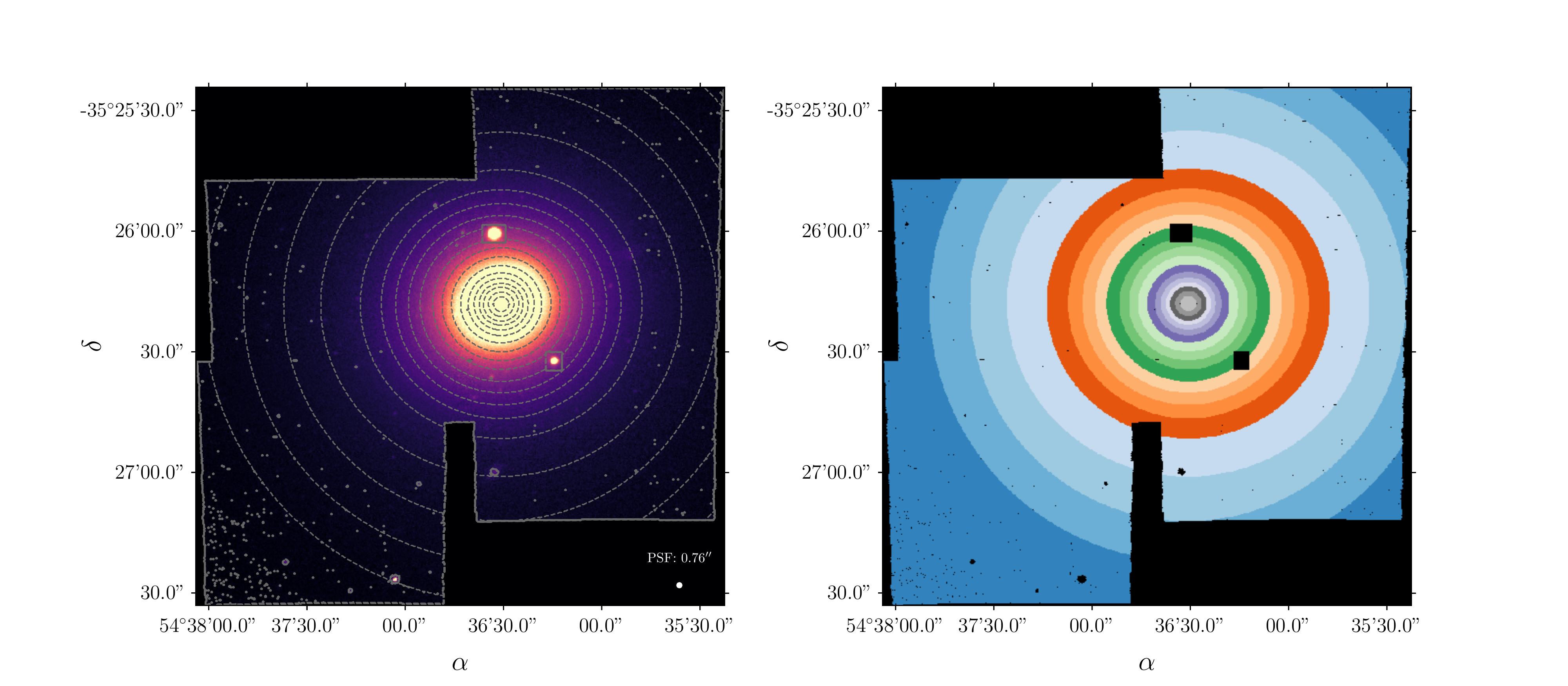}
\caption{The centre of NGC\,1399, divided into eighteen annular bins. The annuli extend to 1.3 times the effective radius.}
\label{fig:annuli}
\end{center}
\end{figure*}

NGC\,1399 was observed on the 2\textsuperscript{nd} and 14\textsuperscript{th} of October 2014 with the Multi Unit Spectroscopic Explorer (MUSE) at the European Southern Observatory (ESO), Paranal, for programme 094.B-0903(A) (PI; Zieleniewski). MUSE combines 24 individual Integral Field Units (IFUs) to give a 1 $\times$ 1 arcminute field-of-view with 0.2\arcsec\,$\times$ 0.2\arcsec\,spatial sampling. The wavelength coverage extends from 4750 to 9300\,\AA. We measure the instrumental resolution, which varies from 75\kms\,at 4800\,\AA\,to 35\kms\,at 9000\,\AA, by fitting to night-sky emission lines. This is discussed further in Appendix \ref{app:inst_res}. 

The galaxy was observed with 5 exposures, each lasting 900 seconds. Four observations were arranged to give a simple mosaic of roughly 2 $\times$ 2 arcminutes square, with the final pointing centred on the middle of the galaxy. Night sky observations, of the same integration time and offset in declination by 5$^{\prime}$, were taken regularly throughout the run to be used as first order sky subtraction. 

The data were reduced with the standard MUSE pipeline (version 1.2.1) and \textsc{esorex} software (version 3.12) to perform the standard bias and dark current subtraction, flat field division and wavelength calibration. The final cube was corrected for small illumination variations using observations of the twilight sky.

The white dwarf CD-30 17706 was observed for the purposes of flux calibration and telluric correction. The most important area of telluric contamination of our spectra is around the NaI absorption line at $\lambda_{\mathrm{rest}}=8190$\,\AA, since the negligible redshift of NGC\,1399 places it is in the middle of the telluric absorption region between 8125 and 8346\,\AA. We carefully scaled the full telluric spectrum to ensure a good removal around this line, and propagated the errors from this process to the final error spectrum for each annulus. After this process, some residual contamination remained between 7160-7330 and 7590-7704\,\AA, but since they did not fall near any important spectral features we chose to simply mask these regions from future analysis.

Subtraction of night-sky emission lines was performed using the dedicated sky observations, on a frame-by-frame basis. We found, however, that significant residuals remained towards the red end of the spectrum, especially in the outer bins. We therefore used the observations to create one-dimensional sky spectra around important telluric emission lines (or families of lines), which are scaled and subtracted during the fitting process itself. See Section \ref{sec:SkySub} for details.

\section{Analysis}

\label{sec:spectral_fitting}

\subsection{Annular bins}
\label{sec:binning}

We bin the datacube into eighteen concentric annuli, masking out two bright objects. NGC\,1399 also hosts a extensive globular cluster system \citep[e.g.][]{2003AJ....125.1908D, 2010A&A...513A..52S}, which may contribute non-negligible amounts of flux to the outer bins. To account for these, we subtract a two component Sersic fit to the light profile of the galaxy and mask any bright residuals before extracting and combining the individual spectra. An image of the collapsed datacube, the annuli and the pixels which have been masked is shown in Figure \ref{fig:annuli}. 

A number of recent studies have found a spatial variation in the low-mass IMF between the galaxy centre and one effective radius (\Reff), the major axis length of an elliptical aperture which encloses half the galaxy light. NGC\,1399, being a cD galaxy \citep{2013AJ....146..160R}, resides in a large, extended stellar halo, and its light profile is most often fit with an inner S\'ersic profile and one or more outer components \citep[e.g. recently][]{2016ApJ...820...42I, 2016ApJ...818...47S, 2017A&A...603A..38S}. In order to compare to a number of other galaxies with resolved IMF measurements (which are not type cD ellipticals), we use the effective radius of the S\'ersic component in our subsequent analysis. This has been measured to be 40\arcsec\,\citep{1991rc3..book.....D, 1988ApJS...68..173D}, meaning that our bins extend to 1.3\,\Reff. 

\begin{figure}
\begin{center}
\includegraphics[width=\linewidth]{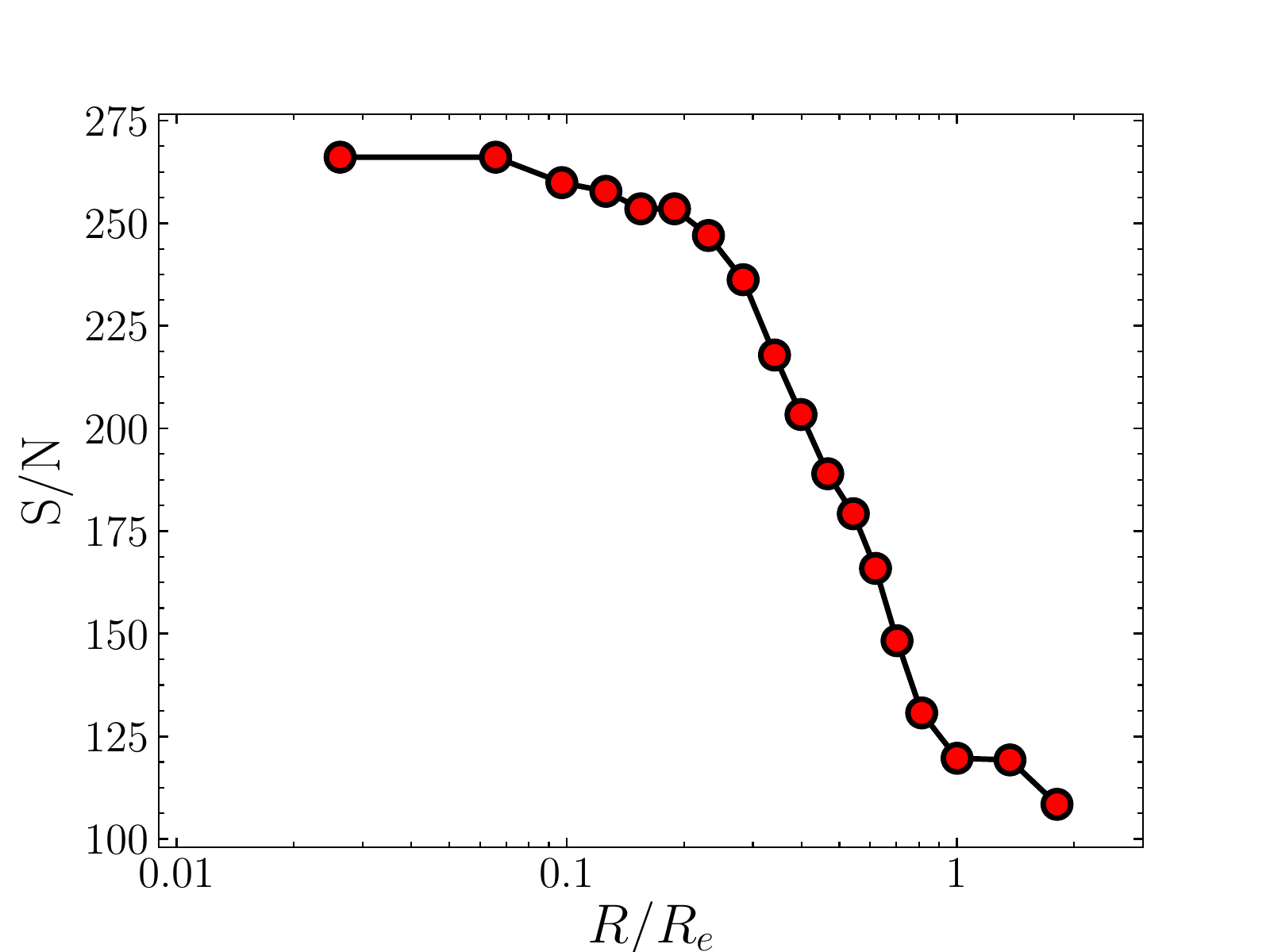}
\caption{True signal-to-noise ratio of the spectrum in each annular bin. These values incorporate the inflation of the error-bars according to the $\beta$ parameter discussed in Section \ref{sec:spectral_fitting}.}
\label{fig:SN}
\end{center}
\end{figure}

Formally, the annular bins each have a signal-to-noise (SN) ratio of well over 1500. However, similarly to the MUSE study of M87 in S18, we find that variation in the residuals around the best fitting template are much greater than one would expect from such a high SN, implying that the true SN ratio of each spectrum is well below this value. S18 attribute this discrepancy to correlations between adjacent pixels introduced during the MUSE data reduction progress, or small instrumental variations between MUSE's 24 integral field units. To account for this under-estimation of the true error in each pixel, we include a parameter in the spectral fitting process to inflate the error-bars appropriately (see Section \ref{sec:spectral_fitting}). The true SN of each spectrum, as found from the ratio of the flux in each spectrum to these rescaled error bars, is shown in Figure \ref{fig:SN}.

\subsection{Stellar Population Synthesis Models}
\label{sec:SPS_models}

In order to make inferences about the stellar population in NGC\,1399, we fit stellar population synthesis (SPS) models to the spectra in each of the eighteen annular bins. We use the stellar template library of \citet{2018ApJ...854..139C}, which model variation in the low-mass IMF, stellar age from 1-13.5 Gyr and metallicity from -1.5 to +0.2 dex across the optical to NIR wavelength range (0.37 - 2.4$\mu m$).  Variation in the individual abundances of 19 elements are are also modelled. 

The models use the MIST stellar isochrones \citep{2016ApJ...823..102C, 2016ApJS..222....8D} and stellar spectra from the MILES \citep{2006MNRAS.371..703S, 2011A&A...532A..95F} and Extended-IRTF \citep{2017ApJS..230...23V} libraries. These library stars are interpolated in the parameters $T_{\mathrm{eff}}$, $\log g$ and metallicity in order to provide empirical spectra for 
each point on the appropriate isochrone. A review of stellar population synthesis techniques can be found in \cite{2013ARA&A..51..393C}.

A key ingredient in the SPS process is the assumed IMF. For these models, the IMF above 1\,\Msun\,is a power law fixed at the Salpeter value, $x=2.3$ \citep{1955ApJ...121..161S}. Below this mass, the power law is broken into two components: the slope between $0.08$\,\Msun$<x<0.5$\,\Msun\,and the slope between $0.5$\,\Msun$<x<1.0$\,\Msun\,(as described in equation 1). These IMF slopes, $x_1$ and $x_2$ respectively, can vary independently and take values between 0.5 and 3.5. 

Variations in individual elemental abundances are handled by using response functions (RFs) to correct the base model at a given age, metallicity and IMF. These RFs are based on theoretical stellar atmosphere calculations which include a comprehensive set of molecular and atomic transitions. RFs are available for spectra with abundance variations between -0.3 and +0.3 dex for a variety of common elements, with enhanced [Na/H] abundances available up to +1.0 dex. These RFs assume a form of the IMF, since varying the IMF and self-consistently computing an elemental RF is, at present, computationally infeasible. All analysis in this work uses RFs which assume a Kroupa low-mass IMF (i.e in equation 1 $x_1=1.3; x_2=x_3=2.3$), although our conclusions are unchanged if, instead, we use the RFs which assume a Salpeter low-mass IMF ($x_1=x_2=x_3=2.3$). 

A full discussion of the models is presented in \cite{2018ApJ...854..139C}. 

\subsection{Spectral Fitting}

We fit the models to the data using the \textsc{python} package \texttt{PyStaff} (Python Stellar Absorption Feature Fitting), which is publicly available at \url{https://github.com/samvaughan/PyStaff}. Below, we outline the steps undertaken during the fitting process.

To ensure that there is a constant velocity difference between pixels which are adjacent in wavelength, we first rebin the data and templates to have an equal sampling in $\ln \lambda$ instead of $\lambda$ \citep[see e.g][]{2017MNRAS.466..798C}. We then linearly interpolate the SPS models in each dimension, allowing us to make a model at an arbitrary set of parameter values. Free parameters in the model which we choose not to vary are fixed at their initial values. This template is then convolved with a line-of-sight velocity distribution (LOSVD), which, in this case, is simply a gaussian function of width $\sigma$ and velocity $v_{\mathrm{syst}}$. We correct for different continuum shapes between the model spectra and the data by fitting multiplicative legendre polynomials to the ratio of the template and the data. The order of this polynomial is 8, 10, 10 and 13 respectively for each of the four regions; we have found negligible change to our results by slightly increasing or decreasing these values. The convolved, continuum matched template at arbitrary parameter values is then compared to the data.

Instead of fitting models to the entire spectral range, we split the process into four sections: 4800-5600\,\AA, 5600-6600\,\AA, 6600-7500\,\AA\,and 7700-9000\,\AA\footnote{The small gap between 7500\,\AA\,and 7700\,\AA\,contains the residuals of a deep telluric absorption feature.}. These sections are then fit simultaneously. The spectral fitting algorithm has 30 free parameters, which are: 

\begin{itemize}
\item Recession velocity ($v_{\mathrm{syst}}$) and velocity dispersion ($\sigma$)
\item Stellar age and metallicity [Z/H]
\item the low mass IMF slopes $x_1$ and $x_2$ (see equation 1)
\item Individual abundance variations in all 19 individual element abundances available in the SPS models, including Na, Fe, Ca, Mg and Ti\footnote{The full list of elements is: Na, Ca, Fe, C, N, Ti, Mg, Si, Ba, O, Cr, Mn, Ni, Co, Eu, Sr, K, V, and Cu}.
\item 4 nuisance parameters regarding the sky subtraction
\item a parameter, $\beta$, to rescale the size of the error bars (see section \ref{sec:binning})
\end{itemize}

We use the affine-invariant Markov-chain Monte-Carlo (MCMC) code \texttt{emcee} \citep{2013PASP..125..306F} to explore this parameter space for each of the spectra. For a given set of parameters, $\theta$, we create a model, $f(\theta)$, and compare to the observed spectrum with values $D_{\lambda}$ and associated errors $\sigma_{\lambda}$ at each wavelength. This corresponds to the following (log)-likelihood function:

\[
\ln p(\theta|D, \sigma)=-\frac{1}{2} \sum_{\lambda} \left[ \frac{(D_{\lambda} - f(\theta)_{\lambda})^{2}}{s_{\lambda}^{2}} + \ln(2\pi s_{\lambda}^{2})\right]
\]

with 

\[
s_{\lambda}^{2}=(1+e^{\beta}) \times \sigma_{\lambda}^{2}
\]
 
for a given value of the nuisance parameter $\beta$. 

We explore this parameter space with 400 walkers. Each walker takes 30,000 steps, of which we discard the first 25,000 as the "burn-in" period. We are therefore left with 2$\times10^{6}$ samples of the posterior in all bins. Each chain is inspected visually for convergence.

\subsection{Sky Subtraction}
\label{sec:SkySub}

First order sky subtraction was performed by subtracting galaxy and sky observations adjacent in time, before the individual data cubes were combined. However, we still found residual sky emission to be prevalent in our observations, especially towards the outskirts of the galaxy. In order to correct for this, we carry out a second sky subtraction during the fitting process \citep[e.g.][]{2009MNRAS.398..561W, 2017MNRAS.465..192Z, 2018MNRAS.475.1073V}. This process is discussed in Appendix \ref{app:sky_sub}

\begin{figure*}
\begin{center}
\includegraphics[width=\linewidth]{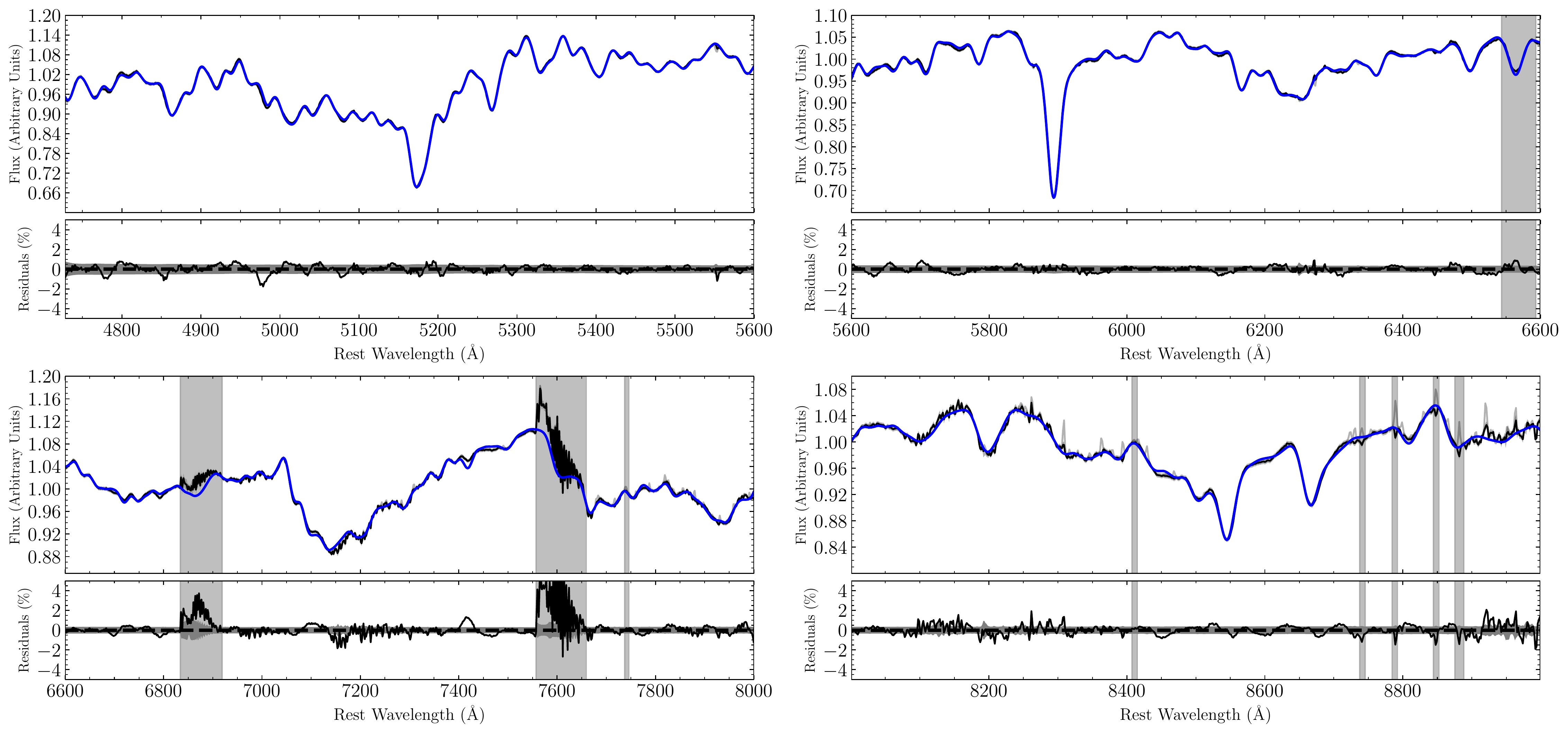}
\caption{The central spectrum of our NGC\,1399 observations (black) compared to the best-fit model (blue).  Residuals between the data and model are shown below each section of the spectrum. Telluric emission lines in the raw spectrum which have been subtracted during the fit are shown in light grey. Shaded regions show areas of residual telluric absorption or emission which have been excluded from the fit.}
\label{fig:fit_centre}
\end{center}
\end{figure*}

\begin{figure*}
\begin{center}
\includegraphics[width=\linewidth]{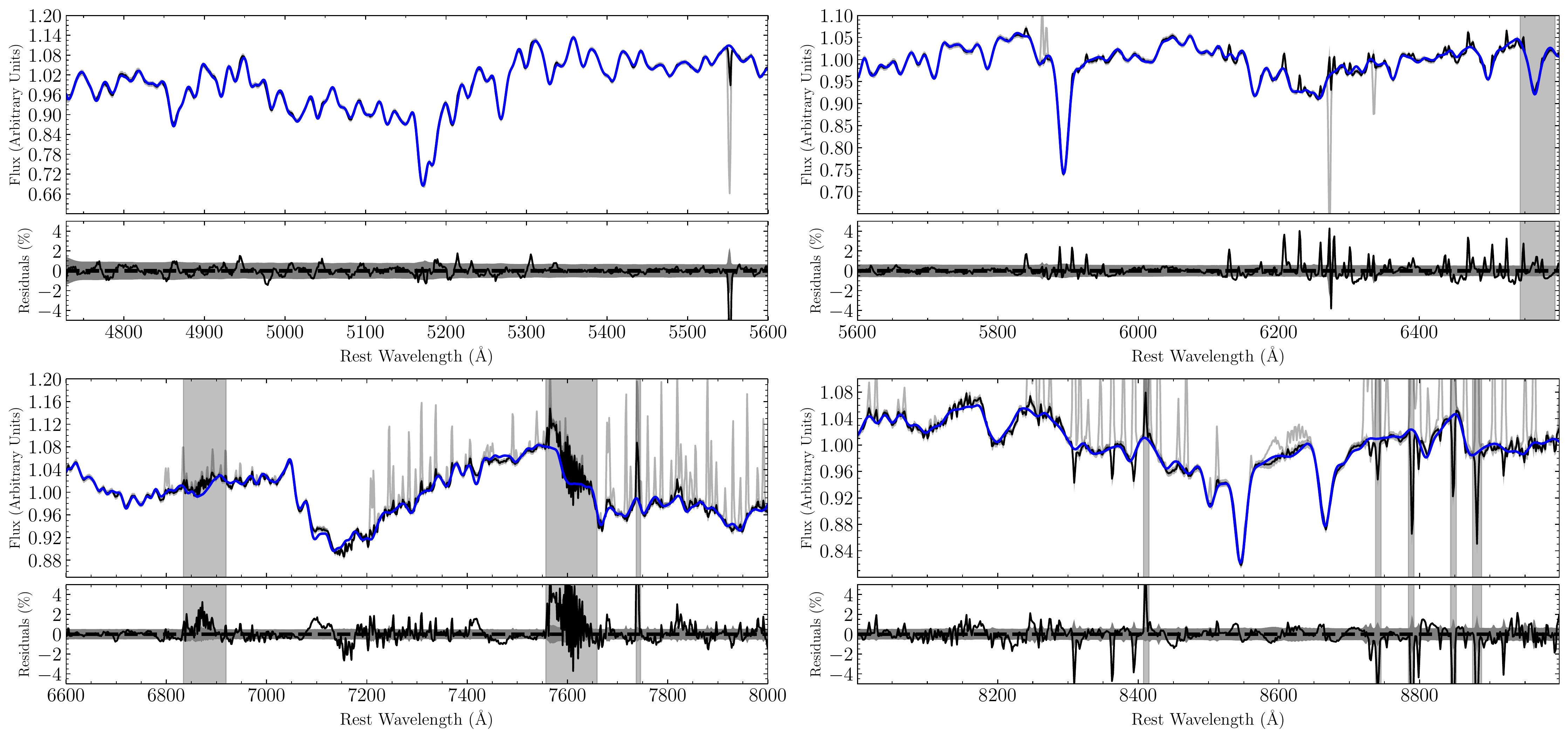}
\caption{As Figure \ref{fig:fit_centre}, but for a spectrum at 20\arcsec\,(0.5 \Reff).}
\label{fig:fit_outer}
\end{center}
\end{figure*}

\begin{figure*}
\begin{center}
\includegraphics[width=\linewidth]{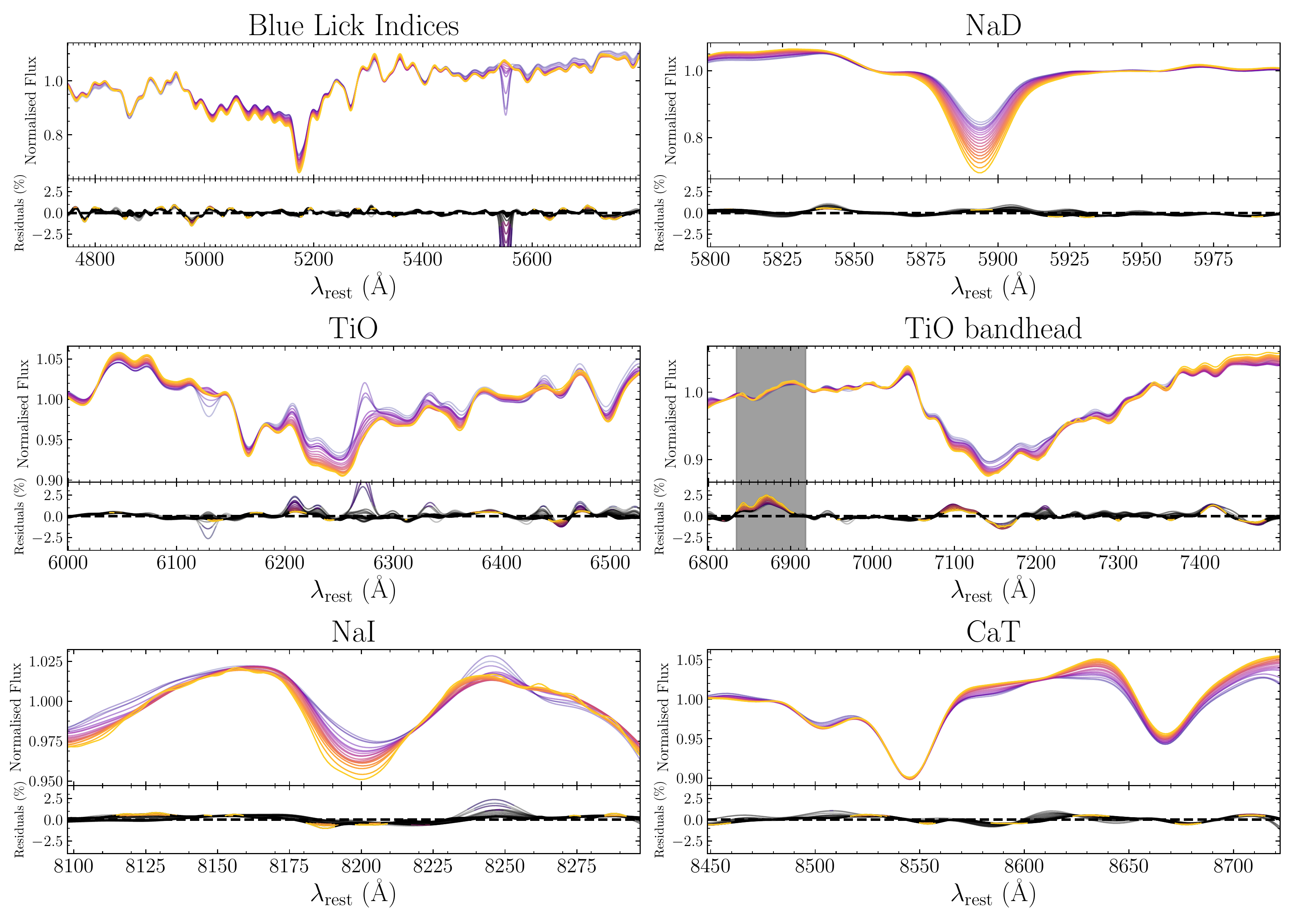}
\caption{Variation in selected areas of the spectra (from the eighteen annular bins) as a function of radius. The data are coloured according to their radius, from yellow (centre) to purple (the outskirts). The spectra have been normalised and convolved to a common velocity dispersion of 350 \kms. The lower panels show the residuals around the best-fit model. Residuals which are within the 1-$\sigma$ uncertainties are black, whilst areas of the spectrum which differ from the model by more than this are plotted in colour. The grey shaded region shows an area excluded from the fit due to residual telluric emission or absorption. Note the different y axis scales for each panel. }
\label{fig:features}
\end{center}
\end{figure*}

\section{Results}
\label{sec:results}
\begin{figure*}
\begin{center}
\includegraphics[width=0.7\linewidth]{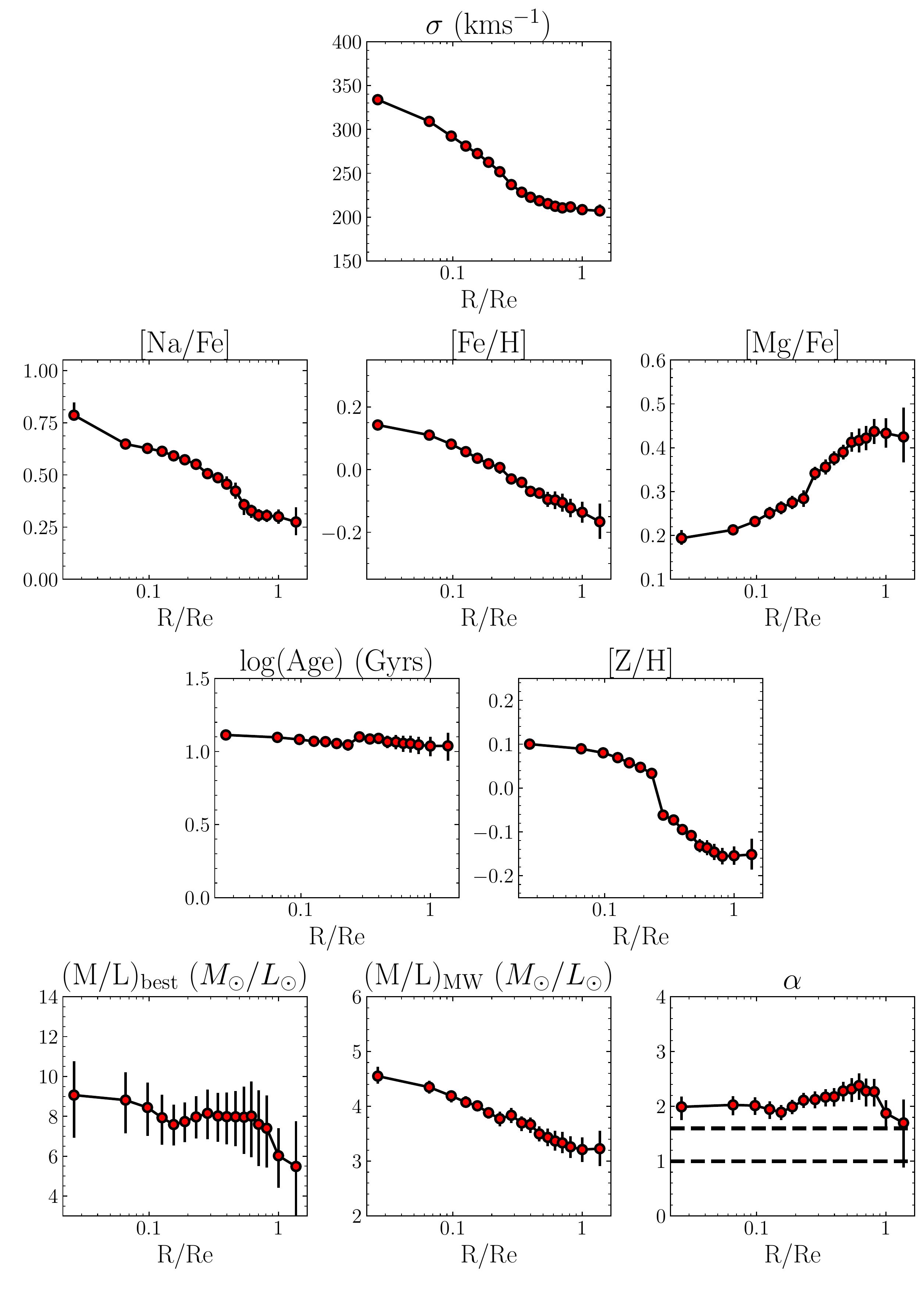}
\caption{Stellar population results as a function of radius from full spectral fitting for NGC\,1399. The main result of this work is the bottom right panel, showing the IMF 'mismatch' parameter $\alpha$ shows no gradient out to $\sim$0.7 \Reff, before falling and becoming marginally consistent with a Milky Way-like IMF at just larger than \Reff.}
\label{fig:results_stellar_pops}
\end{center}
\end{figure*}

The fits to the data are very good, with two examples from the central and outer annular bins shown in Figures \ref{fig:fit_centre} and  \ref{fig:fit_outer}. Similarly to \citet[hereafter vD17]{2017ApJ...841...68V}, we find residuals to be around the sub percent level across the majority of the spectrum. Figure \ref{fig:features} shows the variation as a function of radius for selected spectral features in each of the 18 annular bins, coloured from the centre (yellow) to the outskirts (purple). Also shown are the residuals around the best fitting template as a function of wavelength, with residuals which are greater than the 1-$\sigma$ error shown in colour. 

The strongest radial variation is in the NaD feature at 5890\,\AA. This line is a well known interstellar medium (ISM) absorption line, which complicates its analysis. However, NGC\,1399 has no obvious dust lanes and the line profile is the same as other absorption features in the same spectrum; there are no narrow absorption features which are often associated with ISM contamination \citep[e.g.][]{2004ApJ...610..201S}. We also find no evidence for ISM contamination in unsharp-masked archival B-band HST images of NGC\,1399 (proposal ID 5990; P.I. Grillmair), or in high spatial resolution maps of the NaD line, which we make by measuring the NaD equivalent width in each spaxel in the MUSE observations.  

The NaI 8190 and TiO 6230 features also show modest radial variation, both weakening with increasing radius. The CaT is almost exactly uniform across all radii, however, with only a slight change in the blue and red continuum regions of the third absorption feature evident. Mgb at 5160\,\AA\,weakens slightly with radius, and there are small changes in the assortment of Fe lines in the blue region of the spectrum. Furthermore, the TiO 7053 band head shows a small change between the centre of NGC\,1399 and 0.5 R$_{e}$; the band head is slightly stronger in the centre.

\begin{figure*}
\begin{center}
\includegraphics[width=0.9\linewidth]{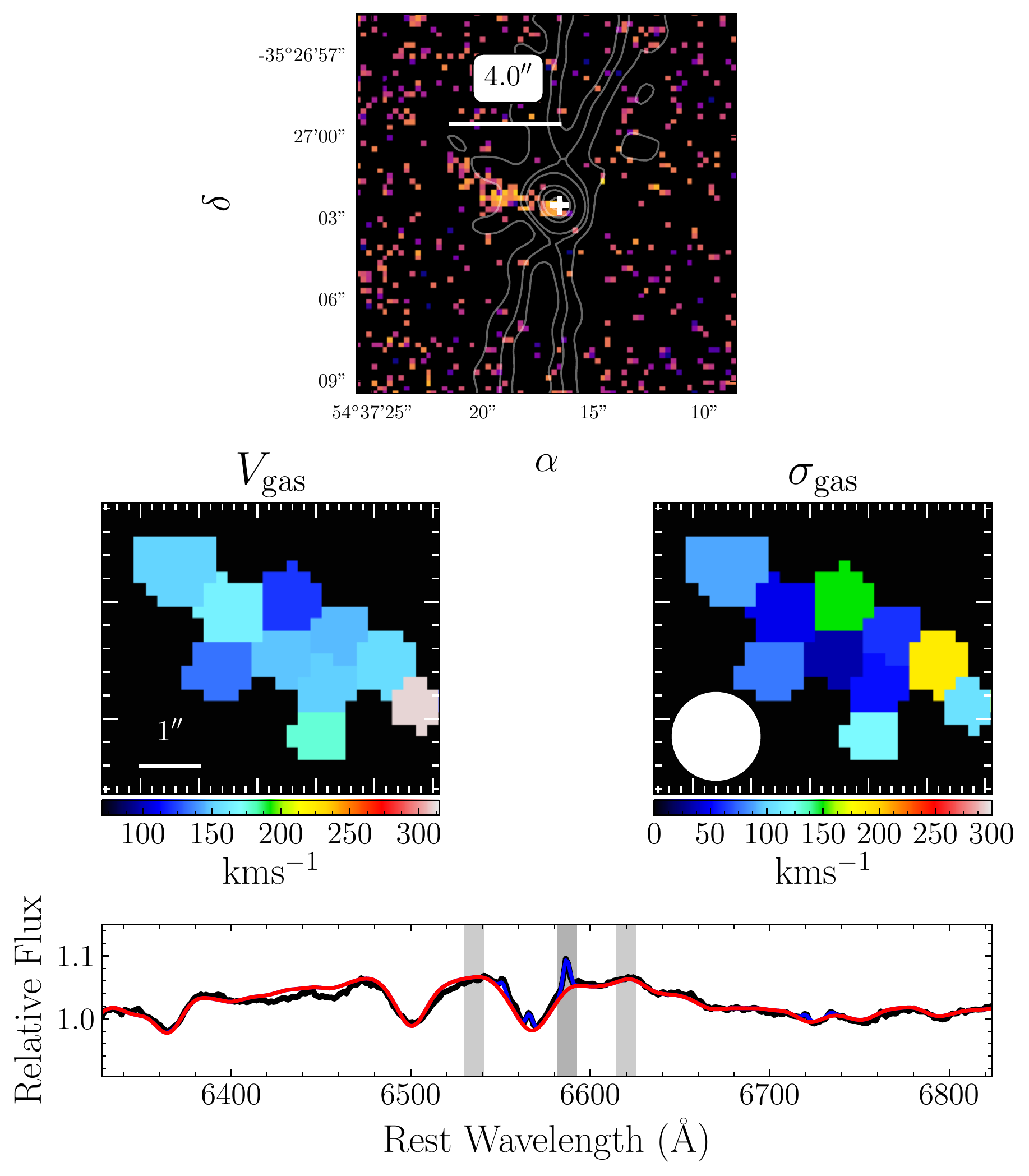}
\caption{The top panel shows a "narrowband" image of the centre of NGC\,1399 from the MUSE observations. We sum flux between 6608.0 and 6619.0\,\AA\,and subtract an average continuum value from between 6556.0 to 6567.0\,\AA\,and 6641.0 and 6652.0\,\AA\,(as shown in the bottom panel). We see an obvious filamentary feature extending 4\arcsec\,from the centre of the galaxy (marked with a white cross). 4.6Ghz radio contours of NGC\`1399's jet, taken from the VLA Archive Survey, are plotted in grey.  NGC\,1399's large-scale radio jets run in the north-south direction. The inset shows a fit to the kinematics of this gas using \texttt{pPXF}, with the seeing radius (of 0.76\arcsec) shown in white. The bottom panel shows a representative spectrum from this feature, with fits to the spectrum shown in red (without gas) and blue (with gas). Fits from \ppxf\,to all ten spaxels in the filament are shown Figure \ref{fig:em_lines}.}
\label{fig:gas_image}
\end{center}
\end{figure*}

\begin{figure}
\begin{center}
\includegraphics[width=\linewidth]{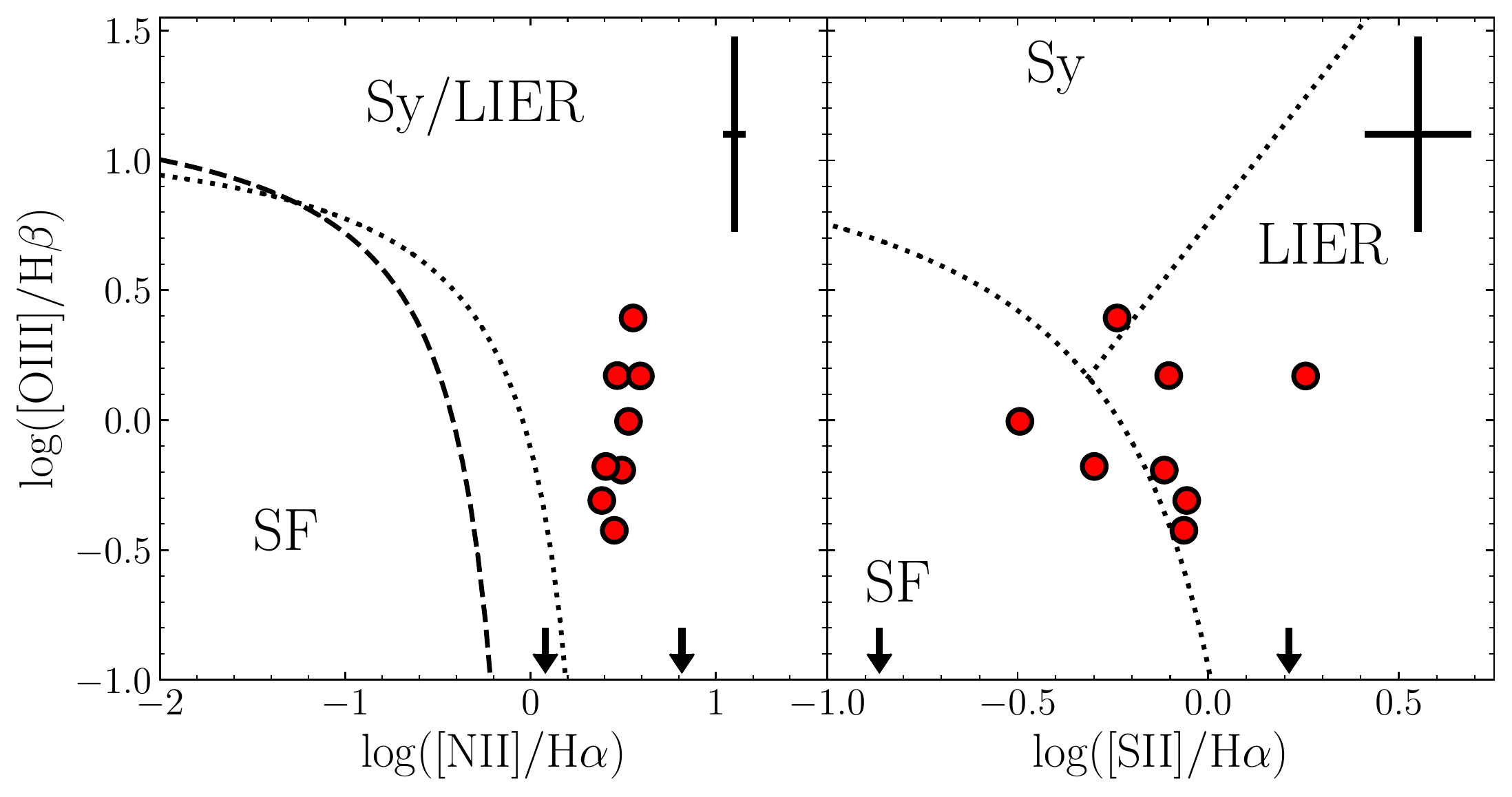}
\caption{Each spaxel in the MUSE observations exhibiting gas emission lines are shown on a BPT diagram. The left plot shows the line ratio [OIII]/H$\beta$ against [NII]/H$\alpha$ whilst the right diagram shows [OIII]/H$\beta$ against [SII]/H$\alpha$. We plot two spaxels with negligible [OIII] emission as upper limits. In the left hand panel, the dashed and dotted lines show the split between starforming and Seyfert/LIER galaxies from \protect\cite{2003MNRAS.346.1055K} and \protect\cite{2006MNRAS.372..961K} respectively. The dashed line in the right panel which separates ionisation caused by star formation, Seyfert radiation and LIERs is also from \protect\cite{2006MNRAS.372..961K}. The majority of spaxels reside in the "LIER" region of the graph. This, combined with the spatial location of the emission lines and the clear evidence for an AGN in NGC\,1399, leads us to conclude that the ionising radiation is from AGN activity.}
\label{fig:BPT}
\end{center}
\end{figure}

\subsection{IMF Measurements}
\label{sec:results}

Figure \ref{fig:results_stellar_pops} shows stellar population parameters in NGC\,1399 derived from spectral fitting. From the top row, the panels show: the velocity dispersion of each bin; the abundances of [Na/Fe], [Fe/H] and [Mg/Fe]; the stellar age and metallicity; and the best fitting M/L ratio (measured in the V-band), the M/L ratio of a galaxy with the best-fitting age and metallicity but with a Milky Way-like IMF, and the IMF "mismatch" parameter $\alpha$, the ratio of these two M/L measurements. A population with a Milky Way IMF will have an $\alpha$ value of 1. 

The steep gradient in [Na/Fe] and the much shallower radial change in [Fe/H] are evident, along with a modest variation in total metallicity [Z/H] of $\sim$0.2 dex. The small gradient in age shows the galaxy is oldest in the middle. The abundance of [Mg/Fe], a proxy for alpha enhancement, shows a positive gradient, qualitatively similar to (although slightly more pronounced than) the results of e.g. \cite{2008MNRAS.385..675S}, \cite{2007MNRAS.378.1507B} and \cite{2015ApJ...807...11G} in massive ellipticals. As noted in vD17, the absolute value of the [Mg/Fe] gradient is very sensitive to the library abundances ([X/Fe] and [Fe/H]) of the stars used to generate the SPS model, although this sensitivity does not effect the relative variations between radial bins or the absolute values of the derived IMF constraints. We refer the reader to \cite{2018ApJ...854..139C} for further discussion of the abundance patterns of the library stars in the models. 

The main results of this work are the bottom panels. The first shows that the inferred M/L ratio in NGC\,1399 displays a negative gradient as a function of radius, decreasing from $9.0^{+ 1.69}_{-1.44}$\,M$_{\odot}$/L$_{\odot}$ in the centre to a value of $5.48^{+4.43}_{-2.27}$\,M$_{\odot}$/L$_{\odot}$ beyond \Reff. The IMF mismatch parameter $\alpha$ is takes the value of $1.97^{+0.19}_{-0.16}$ in the very centre, heavier than a Salpeter IMF (and hence a Milky Way IMF), and is flat out to 0.7 \Reff.  There is tentative evidence that $\alpha$ decreases at radii beyond this, although the error bars on the furthest point are large due to contamination from telluric emission lines. A decrease in $\alpha$ at large radii would be consistent with a number of other studies of nearby ETGs, which find the IMF in the outskirts of similar galaxies to be consistent with that of the Milky Way ($\alpha=1$).

\subsection{Emission Lines}
\label{sec:emission_lines}

Whilst studying the MUSE observations of NGC\,1399, we noticed a small region of low-level ionised gas emission extending $\sim$4\arcsec\,(404 pc at $D_{\mathrm{L}}=21.1$ Mpc) from the centre of the galaxy. This region is larger than the seeing radius of 0.76\arcsec. The presence of emission lines in the centre of NGC\,1399 has been reported before by \cite{2014MNRAS.440.2442R}, although they were unable to comment on the spatial distribution of the ionised gas. 

Further investigation revealed the presence of H$\alpha$ and [N II] $\lambda$6583 emission lines superimposed upon the broad H$\alpha$ absorption feature. The top panel of Figure \ref{fig:gas_image} shows a "narrowband" image created from the MUSE observations, whilst the bottom panel shows a representative spectrum from this region. Also shown are 4.6Ghz radio contours, taken from the NRAO VLA Archive Survey\footnote{http://archive.nrao.edu/nvas/}. The large-scale radio jet in NGC\,1399 is perpendicular to the extended ionised gas emission.

We we use the penalised pixel fitting \ppxf\,method \citep[][]{2004PASP..116..138C, 2017MNRAS.466..798C} to measure the velocity and velocity dispersion of the emission lines. We first aggregate individual spaxels into voronoi-tessellated bins of S/N 300\footnote{Although the true S/N is likely to be lower than this; see Section \ref{sec:binning}}, using the method of \cite{2003MNRAS.342..345C}, and extract a spectrum and error spectrum from each bin. We run \ppxf\,using the MILES stellar library \citep{2006MNRAS.371..703S, 2011A&A...532A..95F}, which contains empirical spectra of nearly 1000 stars in the solar neighbourhood. We also include emission line templates for H$\beta$, H$\alpha$, [SII]$\lambda$6716 and [SII]$\lambda$6731, and the doublets [OIII]$\lambda$5007, [OI]$\lambda$6300 and [NII]$\lambda$6583. We fix the ratio of the [NII]$\lambda$6583 lines to be 3:1, and carefully account for the varying MUSE instrumental resolution at the wavelength of each emission line. This last point is discussed further in Appendix \ref{app:inst_res}. 

We find, however, that the stars from the MILES library give a poor fit to the region just redward of the H$\alpha$ absorption feature, most likely due to the non-solar abundance variations in the centre of NGC\,1399.\footnote{We note that the H$\alpha$ feature was masked during the full spectral fitting process used to measure the IMF, and so such a non-solar abundance does not influence the main conclusions of this work. } To correct for this, we fit each spaxel using just the MILES library and no gas emission lines, in order to measure the residuals between the galaxy and best-fitting template. We then create a response function from the median of these residuals and apply this to each template in the library, before running the fit again with gas emission line templates. This gives a much better fit to the region in question and leads to robust emission line measurements. 

The results from \ppxf\,are shown in the inset panel of Figure \ref{fig:gas_image}, with fits to each spaxel shown in Figure \ref{fig:em_lines}. We find that the emission lines are very narrow, with average velocity dispersion of $65\pm14$\kms. We also see tentative evidence for a velocity gradient across the filament. The best-fit velocity of the spaxel nearest to the centre of NGC\,1399 is 312$\pm$10\kms, whilst the average velocity of all spaxels in the filament is of $160\pm5$\kms. The average stellar velocity in these bins is 194$\pm2$\kms. 

In order to characterise the source of the ionising radiation, in Figure \ref{fig:BPT} we plot each spaxel on a BPT diagram \citep{1981PASP...93....5B} based on the line ratios [OIII]/H$\beta$ against [NII]/H$\alpha$ (left) and [SII]/H$\alpha$ (right). There was no measurable [OIII] emission in two spaxels, and as such these points are plotted as upper limits. To assess the uncertainties of our line ratio measurements, we add noise to a best-fitting template from the \ppxf\,fit and repeat the process of extracting the line fluxes 100 times. A representative error bar for each point is plotted in the 
upper right corner. 

In the left hand panel, we use the classification schemes of \cite{2001ApJ...556..121K} and \cite{2003MNRAS.346.1055K} to differentiate between star-forming regions and areas ionised by Seyfert/LIER (Low Ionisation Emission Region) radiation. Each point lies away from the star-forming region of the diagram, as is expected in the centre of such a passive galaxy. In the right hand panel, using the lines from \cite{2006MNRAS.372..961K},  we classify the majority of the points as ionized by a LIER, rather than a Seyfert-like AGN.

\cite{2010MNRAS.403.1036C} introduce the WHAN diagram (the equivalent width of H$\alpha$ versus [NII]/H$\alpha$) to classify ionisation caused by strong, weak and "false" AGN. They classify galaxies with a equivalent width of H$\alpha<3$\,\AA\,as a "retired" galaxy, i.e. galaxies which have quenched their star formation and are ionised by hot, low-mass evolved stars. The H$\alpha$ emission lines observed in NGC\,1399 are very narrow, well below this $3$\,\AA\,limit, which may suggest that their ionisation is not caused by a low-level weak AGN either. Another possible radiation source of ionising radiation are post asymptotic giant branch (pAGB) stars and hot white dwarfs, which have been shown to be able to account for the ionisation of cold gas observed in ETGs \citep{1994A&A...292...13B,2008MNRAS.391L..29S}.

On the other hand, NGC\,1399 has been shown unambiguously to contain an AGN \citep[e.g.][]{2008MNRAS.383..923S}. Furthermore, \cite{2005ApJ...635..305O} report detection of a transient nuclear point source in the centre of NGC\,1399 in deep far-UV spectra from the Hubble Space Telescope (HST) in January 1999. This point source was detectable in earlier HST images taken in 1996, but not those from 1991 or 1993, and had faded by a factor of 4 by mid-2000. 

The location of the filament of gas, combined with the recent reported activity of the AGN in NGC\,1399, lends weight to the conclusion that the two are related. It is possible that the emission is due to a cloud of cold gas in the process of being accreted onto the central black hole, giving rise to the transient nuclear activity detected in the archival HST images and explaining the difference in velocity between the centre and edge of the filament. In this scenario, the gas cloud's very small equivalent widths may just be a reflection of the (very) low luminosity of the AGN. Follow up high spatial resolution optical spectroscopy would be necessary to investigate this scenario further.

\section{Discussion}

The IMF mismatch parameter, $\alpha$, has been measured as a function of radius using gravity-sensitive spectral indices in a small number of individual galaxies; see, for example \cite{2015MNRAS.447.1033M, 2016MNRAS.457.1468L, 2017ApJ...837..166C}, vD17 and S18. The general consensus is such that the phenomenon of bottom heavy IMFs is confined to the cores of massive ETGs, with measurements in their outskirts more consistent with the IMF in our own galaxy. Broadly speaking, therefore, the measurements in NGC\,1399 are in agreement with this picture. We measure $\alpha$ to be flat out to $\sim$0.7 \Reff\,before a decrease down to a value which is marginally consistent with a Milky-Way IMF at 1.3 \Reff. 

Interestingly, however, there are differences in the shapes of the IMF gradients measured in the works mentioned above. S18 make a comparison between the IMF gradient found in their study of M87 and radial IMF profiles in the massive ETG XSG1 (SDSS J142940.63+002159.0) from \cite{2016MNRAS.457.1468L} and the average IMF measurements in six ETGs from vD17. The IMFs in M87 and XSG1 display very close agreement, with a linear increase in $\alpha$ as a function of $\log R/R_{\mathrm{e}}$ from MW-like at 0.3 \Reff\,to $\alpha\sim2$ in their centres. The vD17 average, on the other hand, has a flatter central profile which reaches $\alpha\sim2.5$, combined with a steeper rise from MW-like at $\sim$0.3 \Reff\,to $\alpha\sim2$ at 0.15 \Reff. Our measurements of NGC\,1399 are more similar to those of vD17, albeit with a bottom-heavy $\alpha$ of >2 measured out to a larger radius (0.7 \Reff) than their average, and our measurements around \Reff\,only marginally consistent with the MW within the (admittedly large) error bar. Individual galaxies in \cite{2017ApJ...841...68V} do show quite significant differences from the binned average, however, likely due to the inherently stochastic nature of galaxy formation, with the NGC\,1399 results being qualitatively very similar to IMF profile in NGC\,1600.

Before too many conclusions are drawn about these differences, it should be noted that  the method of IMF analysis, as well as the population synthesis models used to do the analysis and the form of the IMF itself, differs between S18, \cite{2016MNRAS.457.1468L}, vD17 and this work. S18 and \cite{2016MNRAS.457.1468L} study the strength of selected regions of spectra sensitive to the IMF (as well as age, [Z/H] and other stellar population parameters). They also use a "bimodal" IMF and the MIUSCAT \citep{2012MNRAS.424..157V} and E-MILES \citep{2016MNRAS.463.3409V} models respectively. On the other hand, vD17 and this work analyse the entire spectrum via full spectral fitting, use a two-part power law IMF between 0.08-1\Msun\,(see equation 1) and stellar models from \cite{2018ApJ...854..139C}. A study of the similarities and differences of these technical prescriptions, by applying both methods of analysis to the same object, is sorely needed. 

We do not see a single parameter which correlates well with the $\alpha$ profile. This is in contrast to S18 and \cite{2015ApJ...806L..31M} who see a tight correlation with their bimodal IMF slope $\Gamma_{b}$ and [Z/H]. The tightest correlations with $\alpha$  from vD17 are with radius and [Fe/H], neither of which correlate well with $\alpha$ in NGC\,1399. We do find, in agreement with both studies, that the local stellar velocity dispersion is not a good tracer of the local IMF. 

\cite{2017A&A...603A..38S} present a photometric study of a number of nearby massive ellipticals, including NGC\,1399. They fit the light profile of NGC\,1399 out to 30$^{\prime}$ with two S\'ersic components and an exponential profile. These correspond physically to three components of a massive ETG; a central component of stars, formed in-situ during the galaxy's formation; a component corresponding to remnants accreted from massive progenitors, which sink rapidly via dynamical friction to smaller radii and which have S\'ersic profiles themselves; and a component corresponding to stars accreted from less massive progenitors, which will remain at larger radii and contribute to a galaxy's diffuse outer envelope but add little light to the central regions. 

The break between the central S\'ersic component and the "accreted" S\'ersic component in NGC\,1399 occurs at 30$\pm$6\arcsec\,\citep[][Table 6]{2017A&A...603A..38S}, very close to the radius where the best fit M/L value recovered from our spectral fitting turns over and begins to decrease (26\arcsec). This points towards the break in the best-fit M/L and the turn over in the $\alpha$ parameter being caused by the build up of stars formed in different conditions to the central regions, in agreement with the current paradigm of the inside-out formation of ETGs \citep[e.g.][and references therein]{2013IAUS..295..340N}.

One significant impact of a non-universal IMF in the centres of massive galaxies is on the dynamical measurements of their black holes. In essence, an unresolved population of low-mass stars adds "unseen" stellar mass above that assumed from a MW IMF, which will end up being attributed to a black hole unless a careful study of the stellar M/L is conducted. A dynamical study of the central regions of NGC\,1399 was undertaken by \citet[hereafter H06]{2006MNRAS.367....2H}, measuring a black hole of mass $1.2^{+0.5}_{-0.6}\times10^{9}\mathrm{M}_{\odot}$. Assuming that mass followed light, H06 derive a best-fit dynamical V-band M/L ratio of 9$\pm$1, in excellent agreement with the central V-band M/L of $9.0^{+1.69}_{-1.44}$ from in this work. As a consequence, this implies that central dark-matter fraction in NGC\,1399 is negligible and that the black hole mass estimate from H06 is secure.  
 
\subsection{[Na/Fe] abundance}

The steep gradient in [Na/Fe] abundance is striking. Similar observations have been made in other radial studies of absorption lines sensitive to the abundance of Sodium \citep[e.g. a number of recent studies:][]{2017MNRAS.464.3597L, 2017ApJ...841...68V, 2018MNRAS.477.3954P, 2018MNRAS.475.1073V}, although the explanation for these observed abundance gradients is unclear. 

As discussed in section \ref{sec:results}, we rule out contamination from the interstellar medium (ISM) in our spectra being the cause of the large Na abundance. NGC\,1399 has no visible clumps or dust lanes, the profile of the NaD line shows no evidence for the usual sharp absorption features associated with absorption by a cold ISM and we see no evidence for large-scale dust filaments in HST imaging or 0.2\arcsec\,resolution NaD equivalent width maps. We therefore turn to a number of possible explanations based on the formation of Na in massive and intermediate mass stars. 

Firstly, it is important to note that individual stars in galactic globular clusters have been measured with abundances of [Na/Fe] as high as our measurements in the centre of NGC\,1399, $\geq$0.7 dex \citep[e.g.][]{2006A&A...455..271G, 2009A&A...505..117C, 2017A&A...605A..12M}. Whilst large, therefore, the central [Na/Fe] abundance is certainly not immediately unphysical. 

During the lifecycle of a massive star, Na can be injected into the ISM via stellar winds and in type II supernovae (SNe). There is evidence to suggest that the Na yield from a type II SNe is highly metallicity dependent, with super-solar metallicity SNe producing large [Na/Fe] abundances \citep{2004ApJ...608..405C, 2006ApJ...653.1145K}, implying that the deaths of massive stars in the high metallicity central regions of NGC\,1399 could be the cause. This mechanism for producing large [Na/Fe] abundances would lead to a correlation between [Fe/H] and [Na/Fe], as we observe in NGC\,1399 (panels 2 and 3 of Figure \ref{fig:results_stellar_pops}). Such a correlation is not observed, however, in a similar study conducted in \cite{2017MNRAS.464.3597L}.

AGB stars are also producers of Na. Indeed, chemical models which only include Na production in type II SNe have been found to underproduce globular cluster [Na/Fe] abundances by factors of 2-3 \citep[e.g.][]{1995ApJS...98..617T, 2005NuPhA.758..259G}.  In AGB stars $\geq3$\Msun, Na is produced during hot bottom burning by the Ne-Na cycle $^{22}\mathrm{Ne}+p\rightarrow {}^{23}\mathrm{Na} + \gamma$, although the rate of sodium production from this process is highly dependent on the reactions $^{23}\mathrm{Na}+p\rightarrow {}^{24}\mathrm{Mg} + \gamma$ and $^{23}\mathrm{Na}+p\rightarrow {}^{20}\mathrm{Ne} + \alpha$ which destroy Na \cite[e.g.][]{PhysRevC.65.015801}. Whilst these reactions have recently been measured in laboratory experiments \citep{2013PhRvC..88f5806C}, some uncertainty about Na yields from AGB stars remains; a recent study has only been able to match the observed abundance of [Na/O] in globular clusters from models of thermally pulsating AGB by stars significantly reducing these sodium destruction rates \citep{2017MNRAS.465.4817S}. As in type II SNe, there is some evidence that the abundance of Na in the ejecta of AGB stars increases with metallicity \citep{2013MNRAS.431.3642V}, or in "super" AGB stars in the mass range 6.5-9.0 \Msun\,\citep{2014MNRAS.437..195D}. In the Milky Way, AGB stars have been found to have a small effect [Na/Fe] abundance in the disk \citep{2011MNRAS.414.3231K}, but they may be important in the high metallicity environment of the centres of massive ETGs. 

There are also arguments that an increased abundance of [Na/Fe] may reflect the duration of the star-formation, in a similar way to [$\alpha$/Fe]. \cite{1995MNRAS.277..945T} show that type II SNe produce 100 times as much $^{23}$Na as type Ia SNe. In the short, intense bursts of star-formation in which the centres of massive ETGs were thought to have formed, therefore, it may be expected to see enhanced [Na/Fe] abundance ratios in what now make up their central regions. Whether this effect is enough to quantitatively match the values measured here and in other studies, however, remains to be seen. 

A conclusion to this puzzle would require further study of the production of Na in the extreme conditions which are thought to make up the centre of elliptical galaxies; high densities, super solar metallicities, short formation timescales and a more bottom heavy IMF than the Milky Way. 

\section{Conclusions}
\label{sec:conclusions}

We have presented MUSE observations of the central regions of NGC\,1399, the largest elliptical galaxy in the Fornax cluster. Using state-of-the-art stellar population synthesis models, we measure the low-mass IMF as a function of radius in this object, as well as the chemical abundances of 19 elements and a number of stellar population parameters. We find that the radial profile of the IMF is bottom heavy in the very centre of NGC\,1399, with an IMF mismatch parameter $\alpha=1.97^{+0.19}_{-0.16}$. We measure a flat radial profile out to $\sim$0.7 \Reff\,before $\alpha$ drops, becoming marginally consistent with a Milky Way IMF within the error bars just beyond \Reff. The central V-band M/L implied by our IMF determination is in excellent agreement with the dynamical M/L from \cite{2006MNRAS.367....2H}. 

Our IMF measurements in NGC\,1399 are consistent with the results of radial IMF measurements in other massive ETGs by \cite{2015MNRAS.447.1033M, 2016MNRAS.457.1468L, 2017ApJ...837..166C, 2018MNRAS.477.3954P} and \cite{2017ApJ...841...68V}. Interestingly, the radius at which $\alpha$ begins to decrease (26\arcsec) matches the radius where a fit to the light profile in \cite{2017A&A...603A..38S} transitions between an "inner" S\'ersic component and a S\'ersic component the authors attribute to "accreted" stars. NGC\,1399 may offer evidence, therefore, for the scenario in which massive ETGs form from the inside out, with stars in their central regions forming first in more extreme conditions at high redshift whilst stars currently residing at larger radii being amassed in a series of minor mergers. 

We find a large central abundance of sodium, as often measured in elliptical galaxies, as well as a substantial radial gradient in [Na/Fe]. We argue that the broad shape of the NaD line, combined with the lack of visible dust lanes in NGC\,1399 imply this is not due to absorption by the interstellar medium along the line of sight. The cause of this gradient is not clear, although at super-solar metallicities a large abundance of Na may be created by type II supernovae  \citep{2004ApJ...608..405C, 2006ApJ...653.1145K} and/or produced at higher rates in AGB stars during hot bottom burning \citep{2013MNRAS.431.3642V, 2014MNRAS.437..195D}. 

We also report on the measurement of a filament of ionised gas extending $\sim$4\arcsec (404 pc) from the very centre of the galaxy. The presence of emission lines in the centre of NGC\,1399 was reported previously by \cite{2014MNRAS.440.2442R}, although this work is the first to resolve their spatial extent. We use \ppxf\,to fit and characterise this gas, measuring an average velocity of $160\pm5$\kms\,and velocity dispersion of $65\pm14$ \kms. The spaxel closest to the galaxy centre shows a velocity much greater than this ($312\pm10$ \kms), perhaps signalling an infalling/outflowing scenario related to the central black hole. On a BPT diagram the points lie in the "LIER" region, and the equivalent width of H$\alpha$ is well below the 3\,\AA\,limit which \cite{2010MNRAS.403.1036C} use to differentiate "retired" galaxies from true AGN. However, based on the clear evidence for AGN activity in NGC\,1399 \citep[e.g.][]{2008MNRAS.383..923S} and the central location of the emission, we suggest that (very) low level radiation from the nucleus is the source of ionisation.

\section*{Acknowledgements}

This paper made use of the \texttt{astropy} python package \citep{astropy}, as well as the \texttt{matplotlib} plotting software \citep{matplotlib} and the scientific libraries \texttt{numpy} \citep{numpy} and \texttt{scipy} \citep{scipy}. 

SPV would like to thank C. Conroy for making available the SSP models used in this work. Based on observations taken at the ESO Paranal Observatory under programme ID 094.B-0903(A). This work was supported by the Astrophysics at Oxford grants (ST/H002456/1 and ST/K00106X/1) as well as visitors grant (ST/H504862/1) from the UK Science and Technology Facilities Council. SPV is supported by a doctoral studentship supported by STFC grant ST/N504233/1. RLD acknowledges travel and computer grants from Christ Church, Oxford, and support from the Oxford Hintze Centre for Astrophysical Surveys, which is funded through generous support from the Hintze Family Charitable Foundation. SZ was supported by STFC-HARMONI grant ST/J002216/1.

\bibliographystyle{mnras}
\bibliography{NGC1399}

\appendix

\section{Sky Subtraction}
\label{app:sky_sub}
During the fitting process, we subtract a model of the night-sky emission lines derived from dedicated sky-frames taken during the MUSE observations. These are used to make four one dimensional sky spectra, containing the emission lines of [O I], the NaI D $\lambda\lambda$5890, 5896 doublet, a forest of OH lines and the ro-vibrational O$_2$ band  respectively \citep[using identification from][]{1996PASP..108..277O} These are subtracted from the data during the fitting process. The absolute fluxes of these spectra are free parameters in spectral fitting algorithm, allowing each one to be scaled separately before subtraction.  We found that this was necessary since first order sky subtraction lead to under-subtracted emission lines in the red end of the spectrum but over subtracted ones towards the blue. A plot of the three sky spectra, as well as a galaxy spectrum before and after the full spectral fitting process, is shown in the appendix, Figure \ref{fig:skysub}.

\begin{figure*}
\begin{center}
\includegraphics[width=0.95\linewidth]{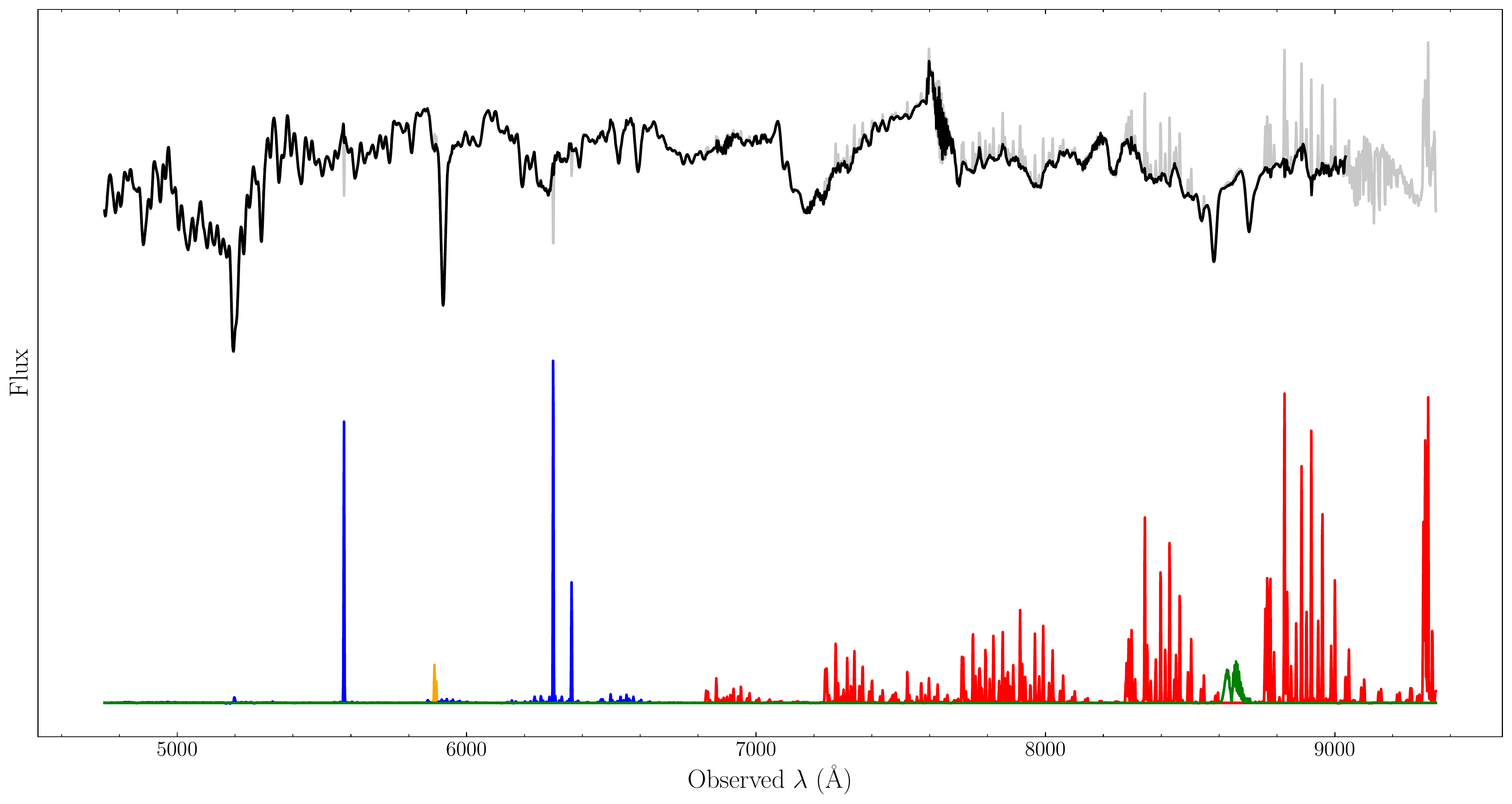}
\caption{A representative galaxy spectrum before (grey) and after (black) the full spectral fitting process, incorporating the sky subtraction technique described in Appendix \ref{app:sky_sub}. The region redward of 9000\,\AA\,is not used in the fit. Below are the four separate one dimensional sky spectra which are used for sky subtraction during the fitting, containing strong [O I] lines (blue), the NaI doublet at 5890\,\AA\,(orange), a forest of OH lines (red) and the O$_{2}$ band at 8650\,\AA\,(green). Note that the grey spectrum, the result of a simple first order sky subtraction, contains under subtracted OH skyline residuals in the red but over subtracted [O I] residuals in the blue.}
\label{fig:skysub}
\end{center}
\end{figure*}

\section{Emission Lines}

\begin{figure*}
\begin{center}
\includegraphics[width=0.95\linewidth]{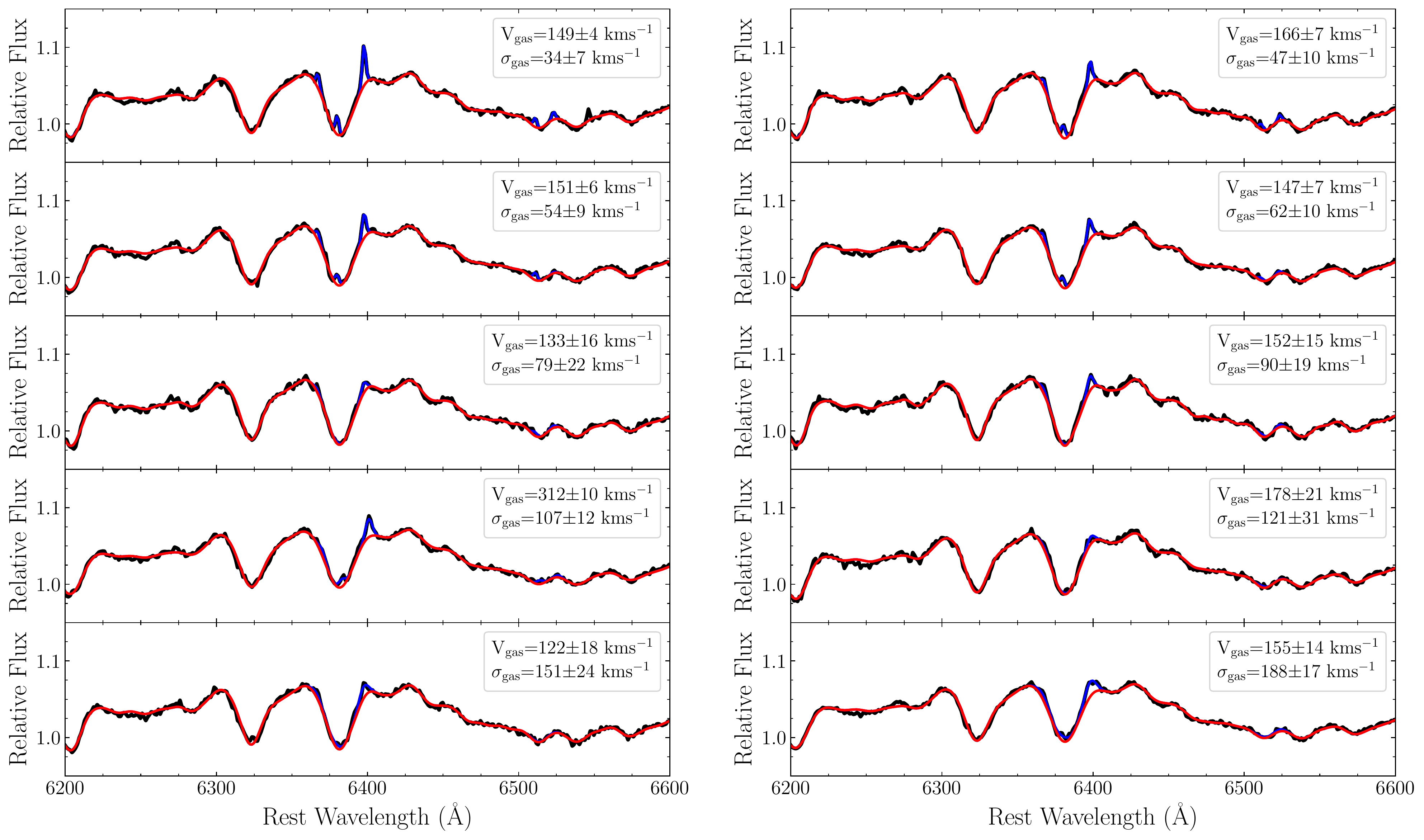}
\caption{Fits to each spaxel in the "filament" of gas shown in Figure \ref{fig:gas_image}. The data are shown in black. We use \ppxf\,to fit the stellar kinematics without gas (red line) and with gas (blue line). Spectra are ordered by velocity dispersion, from narrowest (top left) to broadest (bottom right).}
\label{fig:em_lines}
\end{center}
\end{figure*}

Figure \ref{fig:em_lines} shows the spectra around the H$\alpha$ and [SII] emission lines for each spaxel in Figure \ref{fig:gas_image}. The spectra are ordered by velocity dispersion, from narrowest (top left) to broadest (bottom right).

\section{Instrumental resolution}
\label{app:inst_res}

\begin{figure}
\begin{center}
\includegraphics[width=\linewidth]{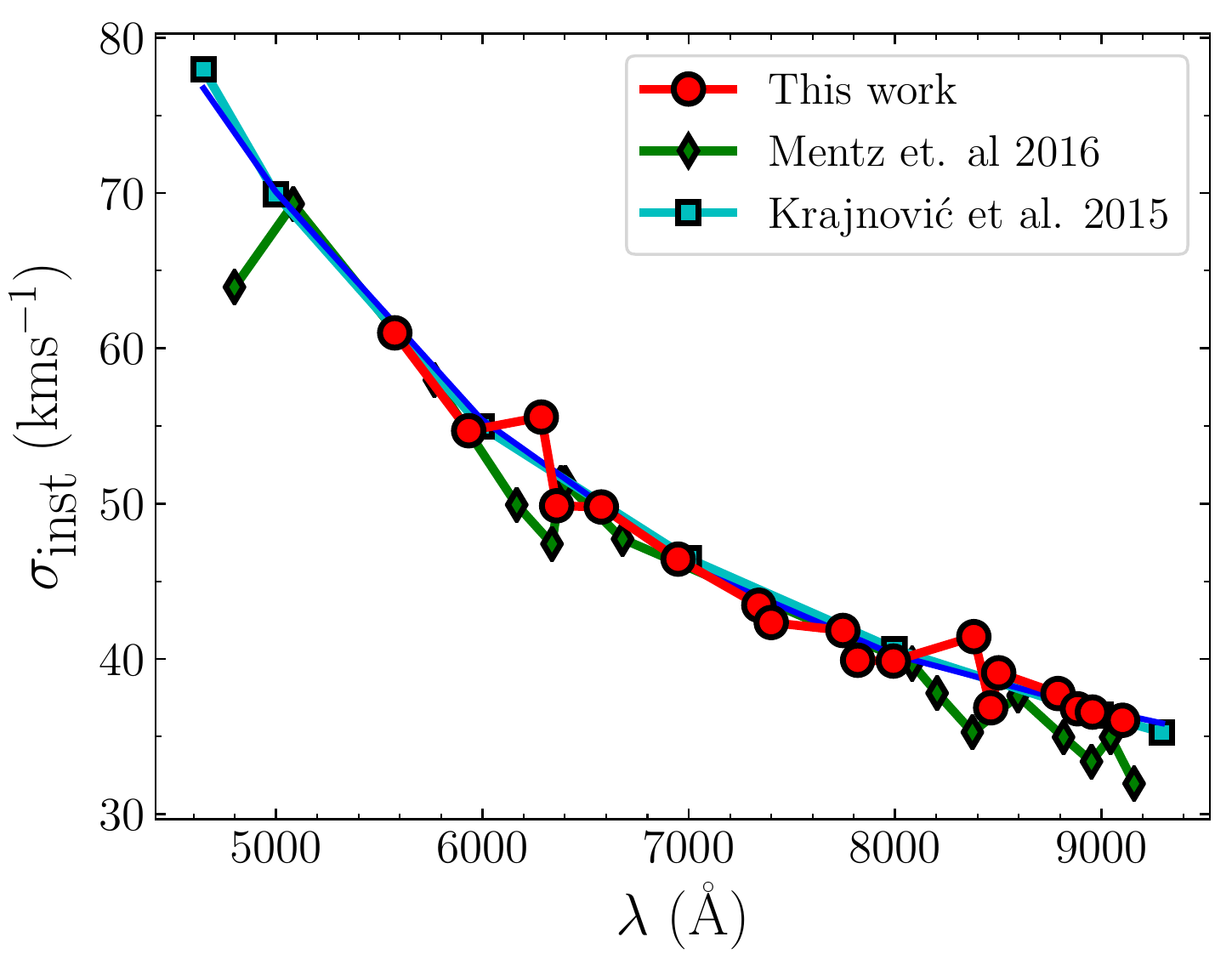}
\caption{MUSE instrumental resolution against wavelength for our data, measured from observations of night sky emission lines, in excellent agreement with measurements from \protect\cite{2016MNRAS.463.2819M} and \protect\cite{2015MNRAS.452....2K}. The blue curve shows a third order polynomial fit to our measurements.}
\label{fig:spectral_resolution}
\end{center}
\end{figure}

In order to characterise the instrumental resolution of MUSE observations as a function of wavelength, we fit gaussians to a number of isolated night-sky emission lines present in a sky frame observation.  Our results are presented in Figure \ref{fig:spectral_resolution}, showing good agreement with other characterisations of the MUSE line-spread function from \cite{2016MNRAS.463.2819M} and \cite{2015MNRAS.452....2K}. We note that the instrumental resolution is negligible compared to the stellar velocity dispersion in all radial bins of NGC\,1399, but is an important consideration when measuring the gas velocity dispersion in Section \ref{sec:emission_lines}. We use our measurements to create emission line templates which closely match the MUSE line-spread function, having an intrinsic width which varies as a function of the emission line central wavelength. This implies that the best-fit dispersion found from \ppxf\,is the true dispersion of the gas, rather than a combination of the gas dispersion and the instrument line-spread function.

\bsp	\label{lastpage}
\end{document}